# Novel theoretical considerations of the coefficient of earth pressure at rest

*To my children, Edor, Geron and Hella may you blossom in wisdom!*
Versions:1(01.11.2021), 2(08.11.2021), 3(03.06.2022), 4(this version 20.08.2023)

Anteneh Biru Tsegaye

Senior Engineer at Norwegian Geotechnical Institute,

anteneh.biru.tsegaye@ngi.no

**Abstract**

One of the well-known and widely applied parameters in soil mechanics is the so-called coefficient of earth pressure at rest (denoted by a $K_0$). Field and laboratory investigations show that this parameter is correlated with the friction angle and the overconsolidation ratio of the soil deposit. For normally consolidated clays, the expression developed by Jaky is considered to hold good. The extension of Jaky's equation for overconsolidated soils is built mainly on empirical observations; and there is lack of a theoretical framework that satisfactorily explains the relationship. In this paper, these relationships are investigated from the nature of energy dissipation of soil aggregates. For the same, the cyclic stress-dilatancy relationship proposed by the author is employed. Then, the stress ratio that maximizes entropy is derived; and this stress ratio is shown to be reasonably close to the coefficient earth pressure ratio proposed by Jaky. An exponential relationship between $K_0$ and OCR is derived; and for a special case, the exponent is shown to be the sine of the critical state friction angle. Finally, some generalizations are given and aspects which need further investigation are highlighted.

**Key words (phrases)**: Soil plasticity, soil mechanics, earth pressure coefficient at rest, $K_0$, lateral earth pressure

# 1 Introduction

A well-known parameter related to an *in situ* stress condition is the so-called coefficient of earth pressure at rest, $K_0$ [1]. Bishop [2] defined the coefficient of earth pressure at rest as the ratio of the lateral to the vertical effective stresses in a soil consolidated under the condition of no lateral deformation, the stresses being principal stresses with no shear stress applied on planes which these stresses act, *i.e.*,

$$K_0 = \sigma_{0h} / \sigma_{0v}, \tag{1}$$

where $\sigma_{0h}$ is the at-rest lateral effective stress assumed equal in all lateral directions and $\sigma_{0v}$ is the at rest vertical effective stress. Note that all stresses in this paper are effective without distinguishing them with a prime.

Measurements and interpretations from intrusive and non-intrusive tests indicated that the coefficient of earth pressure at rest depends on there are several factors [1, 3-7] which is summarized [8] as

$$K_0 = f\left\{\varphi - fric.\,ang., OCR \begin{vmatrix} history \\ age \end{vmatrix}, soil \begin{vmatrix} type \\ properties \\ saturation, mineralogy, cementation, etc. \\ fabric \\ density \end{vmatrix} \right\}, \tag{2}$$

where OCR is overconsolidation ratio given by the ratio of the highest stress experienced by the soil to its current stress. Often, very simple semi-empirical and empirical correlation equations are used to estimate the $K_0$, *e.g.* [1, 4, 9-11]. The most popular relationship that is often used for describing the coefficient of earth pressure at rest for sands and normally consolidated clays is Jaky's [9]. Jaky considered equations of equilibrium to express $\sigma_{0h}$ and $\sigma_{0v}$ in Zone OBC, Figure 1*a*, in terms of the effective unit weight of the soil and the internal friction angle, $\varphi$, which eventually led him to derive the normally consolidated at-rest stress ratio, $K_{0,NC}$, as

$$K_{0,NC} = (1 - \sin\varphi)\frac{1 + \frac{2}{3}\sin\varphi}{1 + \sin\varphi}, \tag{3}$$

which was later simplified to [12]

$$K_{0,NC} = 1 - \sin \varphi. \tag{4}$$

Equation (3) is an exact theoretical solutions for the boundary conditions Jaky considered but Equation (4) seems to be an approximation. Jaky's consideration of the geometry of a heap of sand has less to do with the stress state in a flat terrain (level ground) the $K_0$ is usually associated with and thus one may speculate that the resulting equation is in a reasonable agreement with measurements is perhaps a coincidence. The $K_0$ is also often associated with a one dimensional straining process which has less to do with the mechanism abstracted in the $K_0$ derivation by Jaky [13].

In addition, there is no consistency in literature which friction angle $\varphi$ is considered in the expression for $K_{0,NC}$. Jaky [9] assumed that the friction angle $\varphi$ to be used in the $K_{0,NC}$ is the internal friction angle and that it is equal to the angle of repose, Figure 1b. In some literature, effective angle of shearing resistance is used probably to describe the same thing but not well defined. [6] reported that the maximum the $K_{0,NC}$ value and the minimum void ratio are reached almost simultaneously which implies that the friction angle, $\varphi$ to be used in (4) is density dependent which is less true for the angle of repose. Talesnick [7] further reported that $K_{0,NC}$, and thus $\varphi$, depends not only on density but also on test apparatus, specimen preparation and compaction methods.

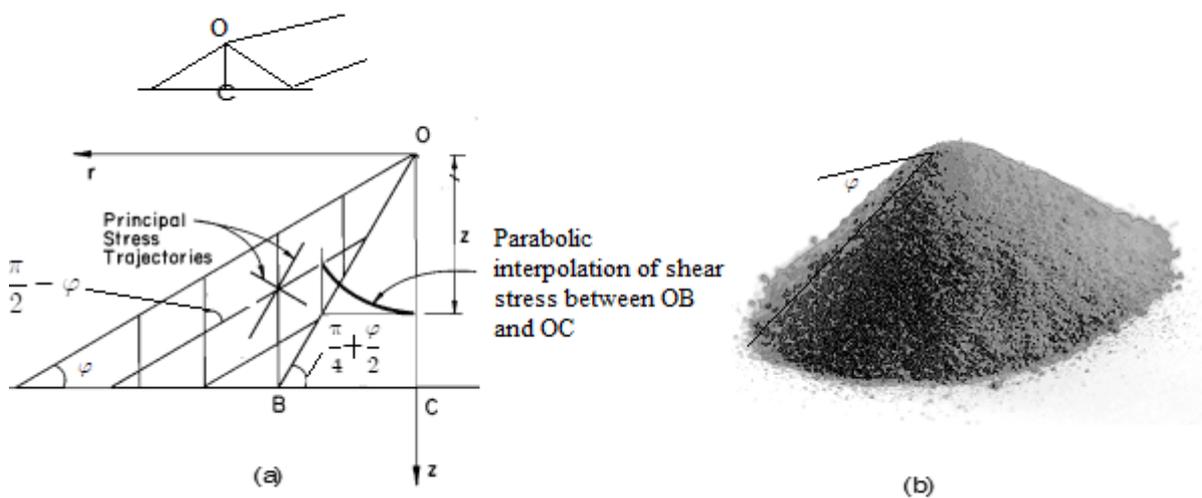

**Figure 1:** (*a*) slip lines in a wage-shaped sand prism, principal stress trajectories and parabolic interpolation of shear stresses between OB and OC, reproduced from [5] and (*b*) definition of angle of repose in a heap of sand.

For sands and normally consolidated clays, Jaky's formula is used as it is or with some minor modifications [1]. For overconsolidated clays, some modifications were proposed, *e.g.*,[1, 4, 10]. The most popular is perhaps the $K_0$ −OCR relationship proposed by Schmidt [10] which goes

$$K_{0,OC} = K_{0,NC} \text{OCR}^x, \qquad (5)$$

where $K_{0,NC}$ is given by Equation (4). Schmertmann [14] suggested that the value of $x$ varies from 0.4 - 0.5. Numerous data have been presented by Mayne and Kulhawy [4] to support the relationship between $K_0$ and OCR; they also arrived at the conclusion that $x = \sin \varphi$. L'Heureux *et al.* [15], based on several field and laboratory investigations arrived at the $K_0 − \text{OCR}$ relationship of the form in Equation (5) for Norwegian clays.

Despite numerous data that are supporting the relationship between $K_0$ and OCR, there is still some skepticism towards the relationship. The unique dependence of the $K_0$ value on OCR has, for instance, been criticized by Jefferies *et al.* [16] who stressed that $K_0$ and OCR are unrelated if the definition in Equation (1) is considered. Accordingly, they suggested that $K_0$ should be regarded an independent variable and measured *in-situ*. Furthermore, Jefferies and Been [17] emphasized that the dominant tendency to rely on empirical correlations to determine *in-situ* properties is undesirable. The author of this paper agrees to this suggestion. However, the repeatedly similar outcome across diverse tests and types of soil may suggest some underlying principle that governs the relationship. There have been some notable attempts that aim at providing a theoretical framework for the justification of the relationship, for instance [18, 19]. In this paper, a novel theoretical framework is established for understanding the relationships between $K_0$, critical state friction angle and OCR which hitherto were in empirical/semiempirical stature. In addition, the effect of dilatancy on the $K_0$ has seldom been directly considered. In fact, Reynolds in his 1985 seminal paper [20] on the experimental illustration of dilatancy envisaged that the recognition of the property of dilatancy would place the earth pressure theory on a true foundation and that it will be useful to show the real reason for the angle of repose. In this paper, the role of dilatancy in the plastic dissipation and further on the at-rest stress ratio in soils is exposed.

Towards this end, the author first establishes the link between plastic dissipation, friction and dilatancy for a plane strain condition. Then, considering Gibbs' conditions of equilibrium, the stress ratio that maximizes the entropy for a given mean effective stress in a plane strain condition is derived and this stress ratio is shown to be reasonably close to Jaky's $K_0$ stress ratio. Furthermore, the $K_0$ created by unloading in shear are found to be proportional to $OCR^x$ in which x is identified with the sine of the critical state friction angle and the constant of proportionality is identified with stress ratio prior to unloading.

Note that:

- Strain rates defined in this paper refer generally to an artificial time increment and can likewise be considered infinitesimal strain increments.
- All stress quantities are effective without distinguishing them with a prime or not necessarily using the adjective "effective".
- Sign convention of soil mechanics is adopted, *i.e.*, compression is positive.
- Rest is defined as a state in which there is neither an external derive nor an internal tendency to changing a given state in to some other.

## 2 Plastic dissipation and stress-dilatancy relationships for loading and unloading in a plane strain condition

Assuming coaxiality between eigen directions of stresses and plastic strain increments, the plastic rate of work in a plane strain condition may be written as

$$D^p = (\sigma_1 + a)\dot{\varepsilon}_1^p + (\sigma_3 + a)\dot{\varepsilon}_3^p, \tag{6}$$

wherein $\sigma_i$ and $\dot{\varepsilon}_i^p$ are principal stress and plastic strain rates respectively ($i = 1$ for major, and $i = 3$ for minor) and $a$ is attraction [21]. Note that strain rates defined in this paper refer generally to an artificial time increment and can likewise be considered infinitesimal strain increments.

Considering Coulomb's shear strength theory in a Mohr-circle, the relationship between principal stress components is written as

$$\sigma_1 + a = N_\varphi (\sigma_3 + a), \tag{7}$$

where $N_\varphi$ is the stress ratio. We also assume that orthogonal plastic strain rates are related as

$$\dot{\varepsilon}_3^p = -N_\psi \dot{\varepsilon}_1^p, \tag{8}$$

where $N_\psi$ is termed here as dilatancy ratio.

The plastic rate of work per unit volume can now be written as

$$D^p = (\sigma_1 + a)\dot{\varepsilon}_1^p d_N, \tag{9}$$

where

$$d_N = 1 - \frac{N_\psi}{N_\varphi}. \tag{10}$$

Considering the hypothesis of complementarity of stress-dilatancy conjugates [22, 23],

$$\delta d_N = -N_\varphi \delta(N_\psi) + N_\psi \delta N_\varphi = 0 \tag{11}$$

yields

$$C_N N_\psi = N_\varphi, \tag{12}$$

where $C_N$ is a 'constant' which may have different values for different modes of shearing. Equation (12) describes a stress-dilatancy relationship, as commonly known, and $N_\varphi$ and $N_\psi$ are referred to as stress-dilatancy conjugates [8]. Note that, although phrased in a more advantageous form, the variation in Equation (11) is equivalent to Rowe's *minimum energy ratio* or *least work* hypothesis.

From Equations (6) and (12), the plastic dissipation is obtained as

$$\mathcal{D}_N^p = (\sigma_1 + a)\dot{\varepsilon}_1^p \left(\frac{C_N - 1}{C_N}\right) \geq 0. \tag{13}$$

Assuming nonnegative plastic dissipation, the inequality

$$C_N = \langle -s \rangle C_N^U + \langle s \rangle C_N^L, \; s = \text{sgn}\left(\dot{\varepsilon}_1^p\right), \; 0 < C_N^U = 1/C_N^L \leq 1, \tag{14}$$

is proposed [8, 23] where $\langle \rangle$ is the Macaulay bracket, the superscripts $L$ and $U$ respectively indicate loading and unloading. Note that $C_N^U$ does not have to be the inverse of the $C_N^L$ except that its value must be positive and less than unity for making sure that the plastic dissipation is nonnegative during unloading. The inverse relationship is just one of the possibilities.

Considering the stress-dilatancy relation in Equation (12) and further assuming $d\sigma_1 \dot{\varepsilon}_1^p + d\sigma_3 \dot{\varepsilon}_3^p = 0$, i.e., associated plasticity, one obtains [22]

$$C_N \frac{d\sigma_1}{d\sigma_3} = N_\varphi, \; N_\psi = \frac{d\sigma_1}{d\sigma_3}. \tag{15}$$

The solution to this differential equation is:

$$\sigma_1 + a = C(\sigma_3 + a)^{\frac{1}{C_N}}. \tag{16}$$

The constant of integration $C$ may then be established by considering a known boundary condition along the curves described by Equation (16). A boundary condition considered here is where along the curve defined by Equation (16) the stress state is isotropic, i.e., $\sigma_1 = \sigma_3$. We call this stress apparent pre-consolidation stress during loading and denote it with $p_c$. We will later differentiate $p_c$ for loading and for unloading. At this point it suffice to write

$$C = (p_c + a)^{\frac{C_N - 1}{C_N}}. \tag{17}$$

Combining Equations (16) and (17), we have:

$$\frac{\sigma_1 + a}{p_c + a} = \left(\frac{\sigma_3 + a}{p_c + a}\right)^{\frac{1}{C_N}}. \tag{18}$$

The stress ratio, $N_\varphi$, can now be given as:

$$N_\varphi = \frac{\sigma_1 + a}{\sigma_3 + a} = \left(\frac{\sigma_3 + a}{p_c + a}\right)^{\frac{1 - C_N}{C_N}}. \tag{19}$$

Or, the plastic potential/yield function can be written as

$$f = \frac{\sigma_1 + a}{p_c + a} - \left(\frac{\sigma_3 + a}{p_c + a}\right)^{\frac{1}{C_N}} = 0. \tag{20}$$

This was previously derived and generalized by the author [22] and is then called Cyclic State Dilatancy (CStaD) plastic flow potential.

The mobilized dilatancy angle may also be written in terms of the mobilized friction angle and the critical state friction angle as follows. Let us consider a Mohr-Coulomb (MC) material where the mobilized stress ratio and the critical state stress ratio are defined respectively as

$$N_\varphi^{MC} = \frac{1+\sin\varphi_m}{1-\sin\varphi_m} \text{ and } C_N^{L,MC} = N_c = \frac{1+\sin\varphi_c}{1-\sin\varphi_c}, \tag{21}$$

in which $\varphi_m$ is the mobilized friction angle, $\varphi_c$ is the critical state friction angle. The mobilized dilatancy angle can then be written as [8]

$$-\sin\psi_m := \frac{1-N_\psi}{1+N_\psi} = \frac{\sin\varphi_m - s\sin\varphi_c}{1 - s\sin\varphi_m \sin\varphi_c}, \tag{22}$$

where $s = 1$ during loading in shear (shear mobilization away from isotropic stress condition) and $s = -1$ during unloading in shear (shear mobilization towards isotropic stress condition) – sign convention of soil mechanics applies, thus contraction is positive. When $s = 1$, the relationship is just the well-known Rowe's [24] stress-dilatancy relationship.

## 3 The coefficient of earth pressure for maximizing entropy

Although the $K_0$ stress state is a geostatic condition, we could envisage that reaching a $K_0$ stress state is an elastoplastic process which involves storage and dissipation of energy. That is, the soil mass must undergo necessary elastoplastic deformations before it reaches the $K_0$ stress state. For the soil mass to maintain the resulting $K_0$ stress ratio, the $K_0$ stress proportion must be a preferred state in terms of equilibrium. In other words, there must not be other neighboring stress proportion that the stress state in the soil tends spontaneously to without it being subjected to external actions that alter its equilibrium state at $K_0$. This can be seen considering Gibbs' equilibrium conditions [25]. Gibbs stated his conditions of equilibrium as follows.

- *For the equilibrium of any isolated system, it is necessary and sufficient that in all possible variations in <u>the state of the system which do not alter its energy</u>, the variation of its entropy shall either vanish or be negative. If ε denotes the energy, and η the entropy of the system, and we use a subscript letter after a variation to indicate a quantity of which the value is not to be varied, the condition of equilibrium may be written*

$$(\delta \eta)_\varepsilon \leq 0.$$

- *For the equilibrium of any isolated system, it is necessary and sufficient that in all possible variations in the <u>state of the system which do not alter its entropy</u>, the variation of its energy shall either vanish or be positive. This condition may be written*

$$(\delta \varepsilon)_\eta \geq 0.$$

Gibbs proved that these two statements are equivalent and therefore satisfying one leads to the satisfying of the other.

Let us consider the first condition of stability and further consider that plastic dissipation is a measure of entropy. The latter holds good for isothermal thermodynamics of a continuum body.

The plastic dissipation for plane strain condition may then be conveniently written as

$$\mathcal{D}^p = (s_m + a)\dot{\gamma}^p \left(\sin \varphi_m + \sin \psi_m\right), \tag{23}$$

where $s_m = \frac{(\sigma_1 + \sigma_3)}{2}$, $\dot{\gamma}^p = |\dot{\varepsilon}_1^p - \dot{\varepsilon}_3^p|$. The sum of the mobilized friction angle and the mobilized dilatancy angle, $\sin \varphi_m + \sin \psi_m$, is assumed to quantify the state of entropy of the system and is denoted by $\eta_{\varphi\psi}$.

We then ask ourselves which combination of the mobilized friction angle and the dilatancy angle maximizes $\eta_{\varphi\psi}$. In other words, we wish to find the condition that maximizes $\eta_{\varphi\psi}$ for an infinitesimal virtual perturbation in terms of plastic shear strain increment (and thus mobilization of friction angle.) This may be investigated by considering the variation

$$\frac{\mathrm{d}\eta_{\varphi\psi}}{\mathrm{d}\sin \varphi_m} = 0, \quad \eta_{\varphi\psi} = \frac{\mathcal{D}^p}{s_m \dot{\gamma}^p} = \sin \varphi_m + \sin \psi_m, \tag{24}$$

which yields [8] (compression positive),

$$\sin \varphi_m = \sin \psi_m = \frac{1 - \sqrt{1 - \sin^2 \varphi_c}}{\sin \varphi_c}. \tag{25}$$

The plastic dissipation for a plane strain deformation mode for a given $s_m$ and at maximum entropy (defined by maximum $\eta_{\varphi\psi}$) may then be written as

$$\begin{aligned} \mathcal{D}_\eta^p &= \frac{2 s_m \dot{\gamma}^p}{\sin \varphi_c} \left(1 - \sqrt{1 - \sin^2 \varphi_c}\right) = 2 s_m \dot{\gamma}^p (1 - \cos \varphi_c)/\sin \varphi_c \\ &= s_m \dot{\gamma}^p \left(1 + \frac{\sin \varphi_c}{4} + \frac{\sin^2 \varphi_c}{8} ...\right) \end{aligned}. \tag{26}$$

Considering the mobilized friction angle given in Equation (25), the corresponding stress ratio can now be readily obtained as

$$\frac{\sigma_3 + a}{\sigma_1 + a} = \frac{1}{\sqrt{N_c}} = \sqrt{\frac{1 - \sin \varphi_c}{1 + \sin \varphi_c}}. \tag{27}$$

Performing Taylor expansion, we find

$$\sqrt{\frac{1 - \sin \varphi_c}{1 + \sin \varphi_c}} = 1 - \sin \varphi_c + \frac{\sin^2 \varphi_c}{2} - \frac{\sin^3 \varphi_c}{2} + \frac{3 \sin^4 \varphi_c}{8} + ..., \tag{28}$$

which implies that $1 - \sin \varphi_c$ is the first order approximation of $\sqrt{(1 - \sin \varphi_c)/(1 + \sin \varphi_c)}$. This stress ratio is reasonably close to Jaky's formula in Equation (4) at lower friction angles. At higher friction angles, the higher order terms seem to gain some weight and hence the value of $\sqrt{(1 - \sin \varphi_c)/(1 + \sin \varphi_c)}$ tends to deviate from Jaky's formula. Visual comparison is given in Figure 2. In addition, plots of normalized plastic dissipation, mobilized dilatancy angle and stress ratio are shown in Figure 3. The tendency to maximize entropy, in line with Gibbs' conditions of equilibrium [25], may therefore be the reason why the stress ratio at rest for normally consolidated soil deposits tends to a certain $K_0$ value. However, for higher values of critical state friction angle, Equation (27) gives higher stress ratio than obtained from Jaky's formula (4) and there can be several reasons for that.

Possible non-coaxiality between eigen directions of plastic strain rates and that of stresses, which is a form of inefficiency in the energy dissipation of the system, may be one of them.

Considering the schematics in Figure 4, for the stress state at the maximum dissipation, the following relations hold:

$$\left.\frac{p_{cl0}+a}{\sigma_3+a}\right|_{K_0} = C_N^{\frac{C_N}{2(C_N-1)}}; \quad \left.\frac{p_{cl0}+a}{\sigma_1+a}\right|_{K_0} = C_N^{\frac{1}{2(C_N-1)}} \tag{29}$$

Therefore,

$$N_\varphi = \frac{\sigma_1+a}{\sigma_3+a} = \left(\frac{\sigma_3+a}{p_{cl0}+a}\right)^{\frac{1-C_N}{C_N}} = \sqrt{C_N}, \tag{30}$$

and at this stress state the friction angle is

$$\sin\varphi_m = \frac{\sqrt{C_N}-1}{\sqrt{C_N}+1}. \tag{31}$$

It can also be seen that the plastic strain increments along the line of the maximum entropy is always contractive. With the dilatancy angle given, we can also show that, when slightly perturbed, the ratio of the plastic strain increments and the stress increments are in the proportion:

$$N_\psi = \frac{d\sigma_1}{d\sigma_3} = \frac{1}{\sqrt{C_N}}. \tag{32}$$

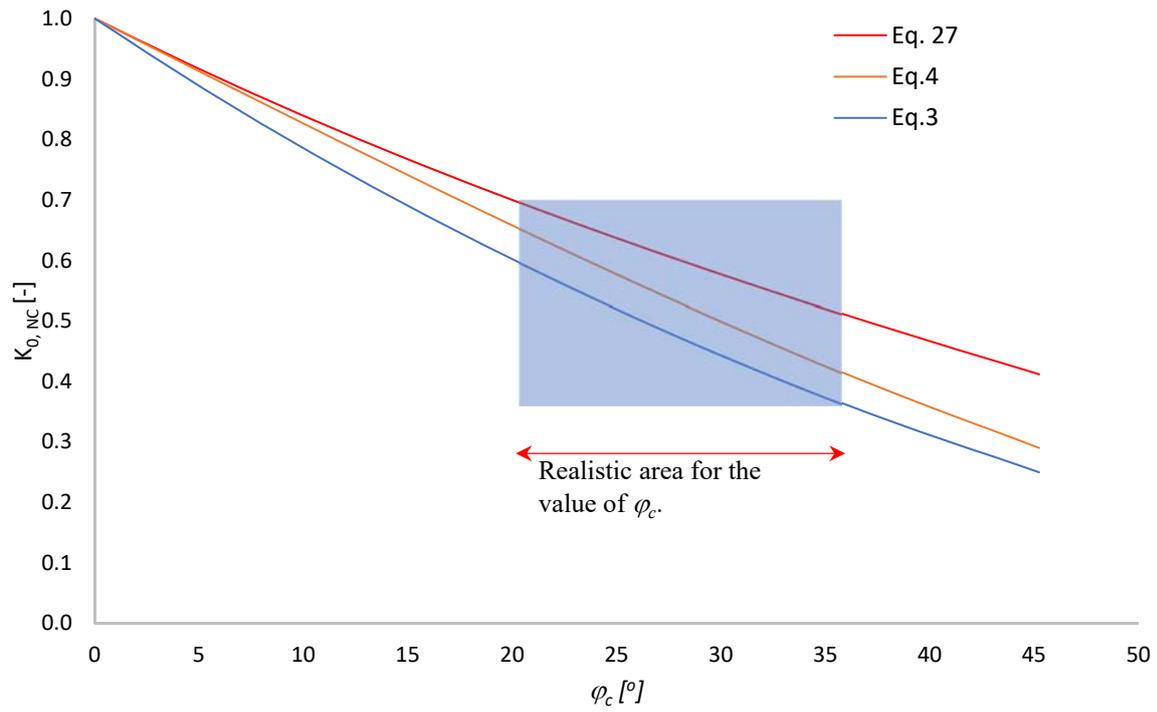

**Figure 2:** Comparison of Equation (27) with Jaky's $K_0$ formulation

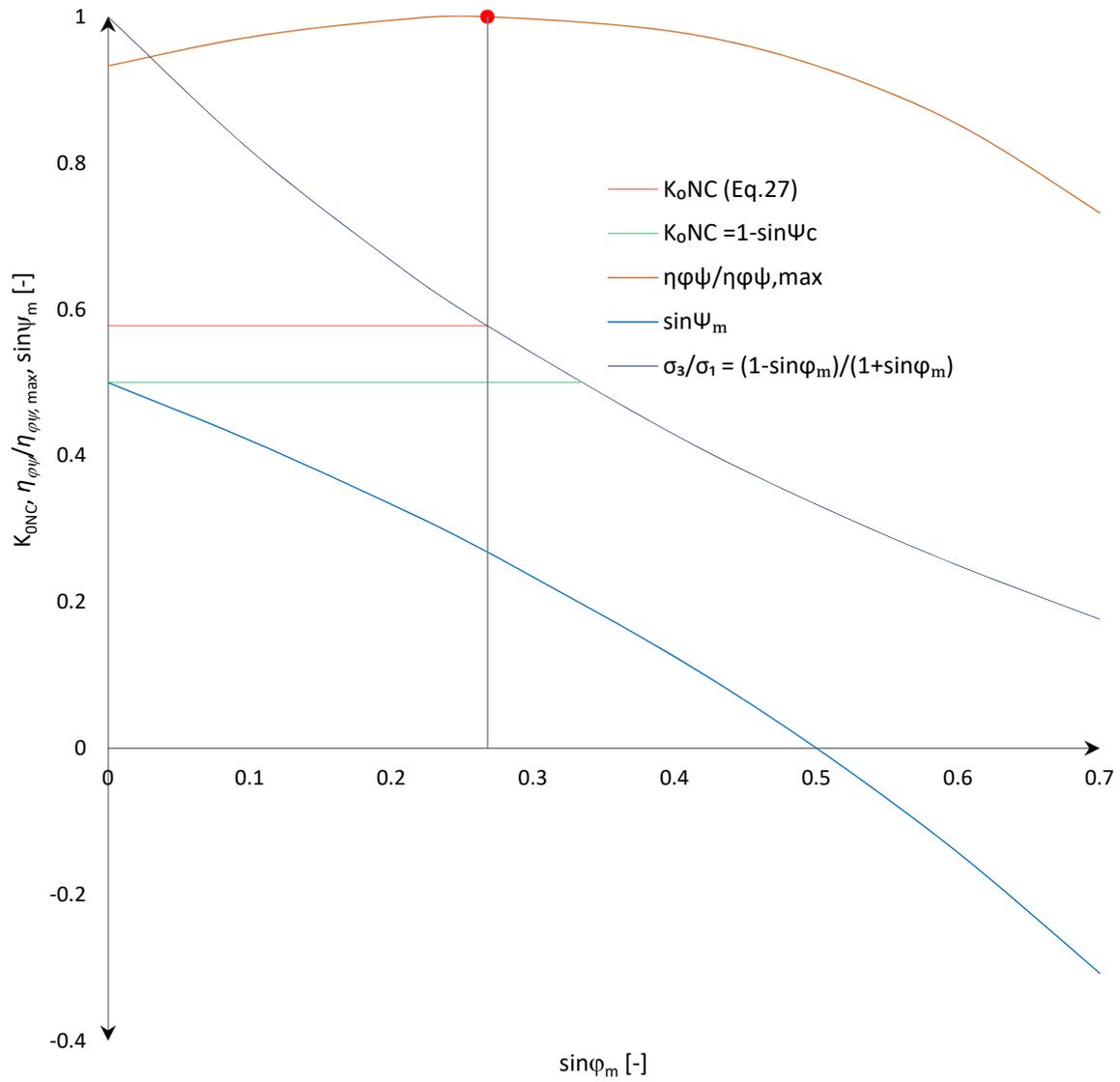

**Figure 3:** Illustration of relationships between mobilized friction angle, mobilized dilatancy angle, normalized plastic dissipation and stress ratio and the $K_{0NC}$ stress ratios according to Jaky's and according to Equation (27) for a critical state friction angle of 30 degrees.

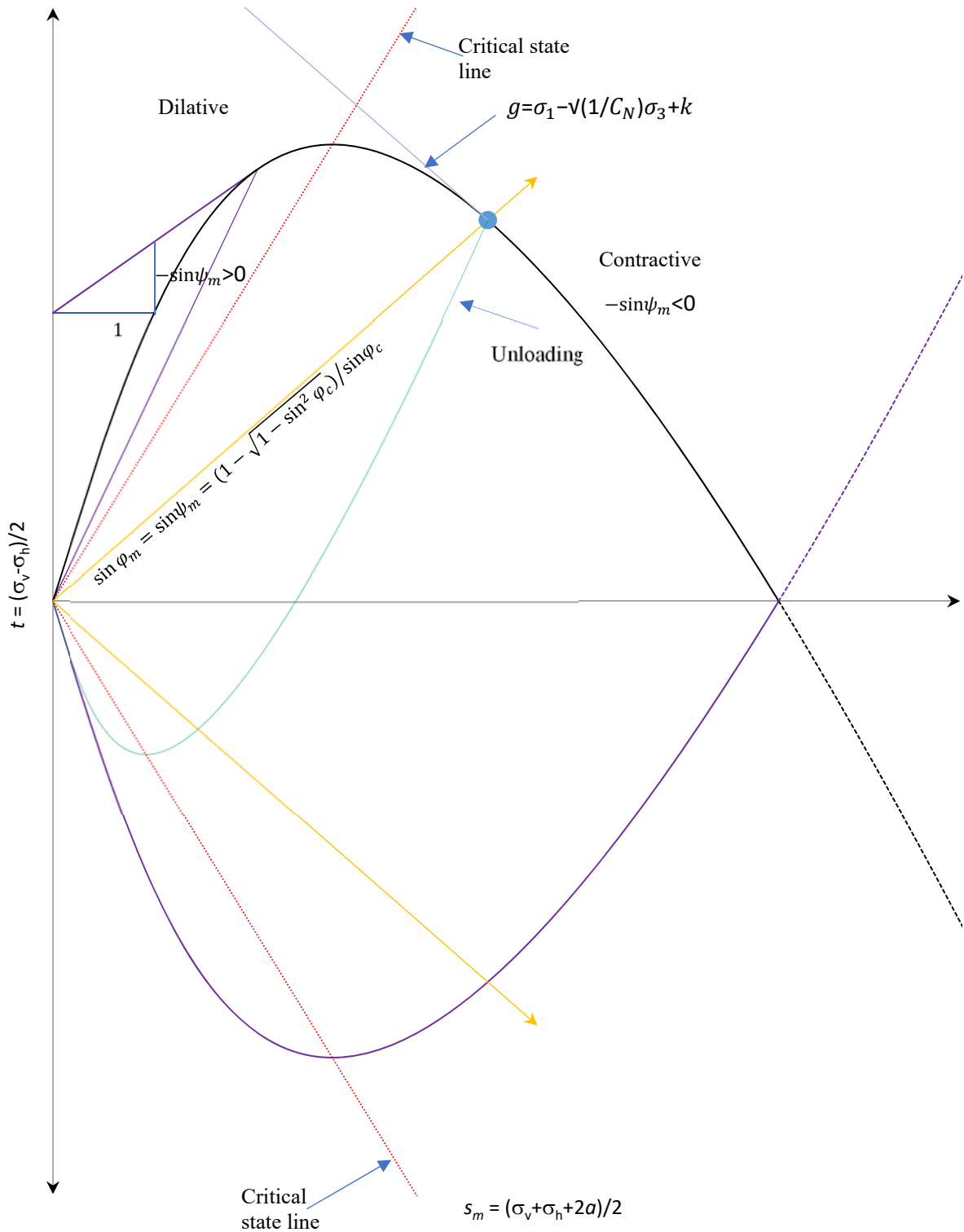

**Figure 4:** Schematics of yield functions during loading and unloading, critical state and line of maximum entropy (defined by maximum $\eta_{\varphi\psi}$) in $s_m$ - $t$ space, $\boldsymbol{\sigma_h}$ and $\boldsymbol{\sigma_v}$ are, respectively, the horizontal and the vertical effective stresses assuming each to be principal stresses.

# 4 Relationship between $K_0$ and unloading induced OCR

We have now shown that a stress ratio at the maximum entropy for a plane strain condition is reasonably close to the coefficient of earth pressure defined by Jaky's equation (4). We will now proceed further with our exposition and show how $K_0$ created by unloading is related to the OCR. For the same, we will consider the schematics in Figure 5 which is a plot of the yield function presented in Section 3 for varying values of apparent pre-consolidation stress. Note that the yield curves are infinitely many and only a few are shown for illustration. The material pays in terms of plastic strains while mobilizing from one stress state to another neighbouring stress state across several yield curves. The yield curve that contains A is assumed for a virgin loading and the material is assumed not to have experienced a stress beyond this curve. For this virgin state, it is assumed that the stress ratio would preferably be at point A for maximizing its entropy for a given mean stress, $s_m$. During unloading from A to B, if not forced, the stress state will mobilize closely following the plastic potential function for unloading, indicated by the green path. This postulate is only an approximation. The stress changes during unloading may mobilize the stress state to a neighbouring yield curve that is described by a different $p_{cu}$ (defined below). This mobilization is assumed to be rather limited while moving towards isotropy, but it may gain significance after crossing the line of isotropy.

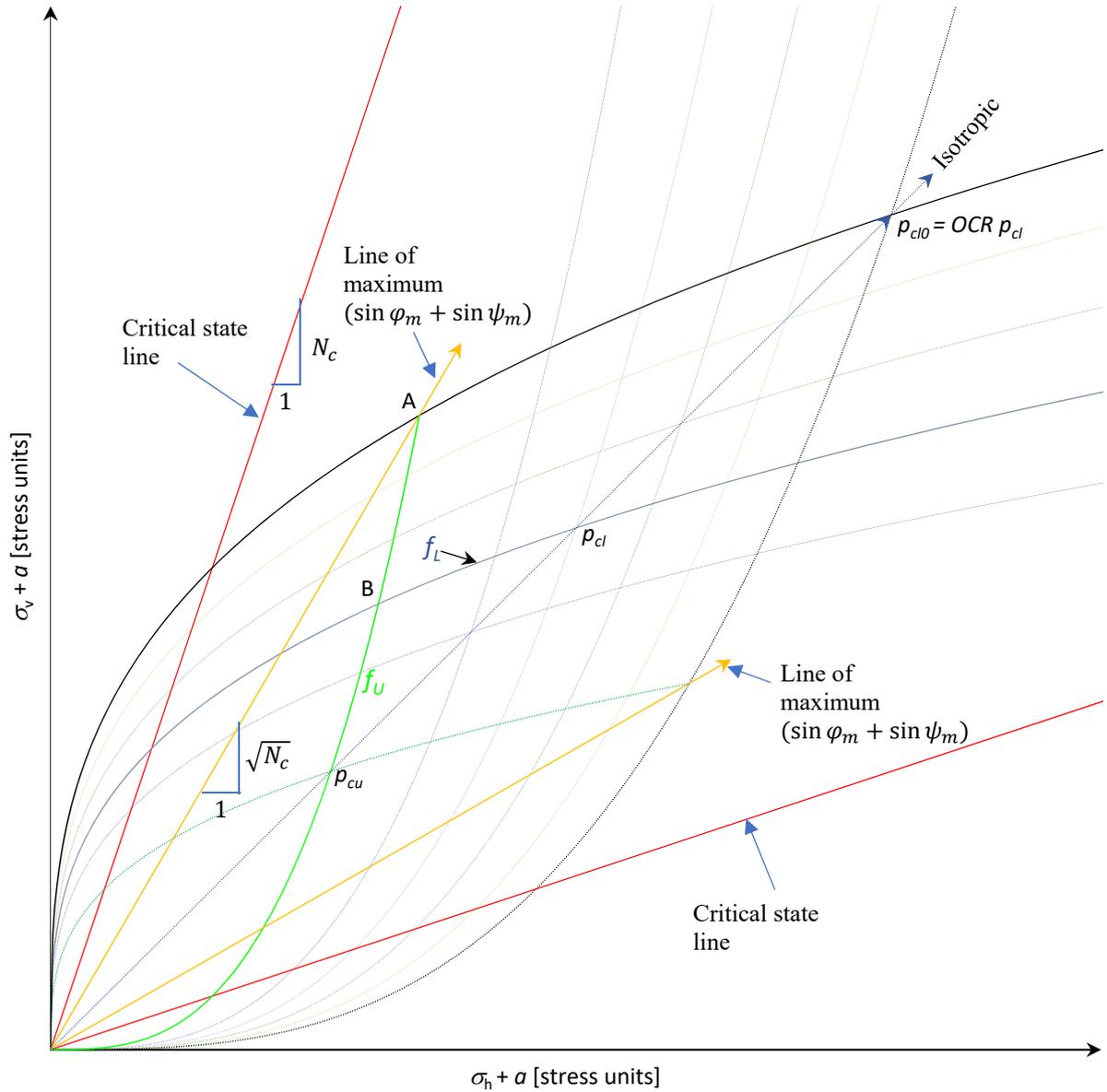

**Figure 5:** Schematics of a set of loading and unloading plastic potential curves, the path of maximum entropy (defined by maximum $\eta_{\varphi\psi}$) and critical state lines, $\sigma_h$ and $\sigma_v$ are, respectively, the horizontal and the vertical effective stresses assuming each to be principal stresses.

The yield function, the stress state obeys during unloading can be written as

$$f_U = \frac{\sigma_1 + a}{p_{cu} + a} - \left(\frac{\sigma_3 + a}{p_{cu} + a}\right)^{N_c} = 0, \qquad (33)$$

in which $p_{cu}$ is where the path intersects the isotropic axis. As we have shown earlier, the stress state at the maximum dissipation is described by

$$\left(\sigma_1 + a\right)\big|_A = \sqrt{N_c}\left(\sigma_3 + a\right)\big|_A. \tag{34}$$

Combining Equations (29) and (34) leads to

$$N_c^{\frac{N_c}{2(1-N_c)}}\left(p_{cl0} + a\right) = \left(\sigma_3 + a\right)\big|_A, \tag{35}$$

and

$$N_c^{\frac{N_c+1}{2(N_c-1)}} = \frac{p_{cl0} + a}{p_{cu} + a}, \quad N_c^{-\frac{N_c+1}{2(N_c-1)}}\left(p_{cl0} + a\right) = p_{cu} + a. \tag{36}$$

For a critical state friction angle between 20 and 36 degrees, which is a realistic range for most soils, the value of the ratio $\frac{p_{cl0}+a}{p_{cu}+a}$ varies between 2.8 and 3.15. The ratio varies between 3.4 and 4.52 when Jaky's equation is used for normally consolidated condition.

Further substituting the relations in Equation (36) into Equation (33) one is led to

$$f_U = \frac{\sigma_1 + a}{p_{cl0} + a} - N_c^{\frac{(N_c+1)}{2}}\left(\frac{\sigma_3 + a}{p_{cl0} + a}\right)^{N_c} = 0. \tag{37}$$

Considering $p_{cl0} + a = \text{OCR}(p_{cl} + a)$ into Equation (37) and further rearranging, one is led to:

$$\text{OCR}^{N_c-1}\left(\sigma_1 + a\right)N_c^{-\frac{(N_c+1)}{2}} = \left(p_{cl} + a\right)^{1-N_c}\left(\sigma_3 + a\right)^{N_c}. \tag{38}$$

Rearranging Equation (20) for the current stress state one obtains

$$\frac{\left(\sigma_1 + a\right)^{\frac{N_c}{N_c-1}}}{\left(\sigma_3 + a\right)^{\frac{1}{N_c-1}}} = p_{cl} + a. \tag{39}$$

Substituting Equation (39) into Equation (38)

$$\text{OCR}^{N_c-1}\left(\sigma_1 + a\right)^{N_c+1} N_c^{-\frac{(N_c+1)}{2}} = \left(\sigma_3 + a\right)^{N_c+1}, \tag{40}$$

which gives

$$\sigma_3 + a = \text{OCR}^{\frac{N_c-1}{N_c+1}} N_c^{-\frac{1}{2}} (\sigma_1 + a). \tag{41}$$

Note that $\frac{N_c-1}{N_c+1} = \sin \varphi_c$, and therefore, the $K_0$ stress state may be stated as:

$$\sigma_h + a = K_{0,NC} \text{OCR}^{\sin \varphi_c} (\sigma_v + a); \; K_{0,NC} = N_c^{-\frac{1}{2}}, \tag{42}$$

where $\sigma_h$ and $\sigma_v$ are, respectively, the horizontal and the vertical effective stresses assuming each to be principal stresses.

Equation (42) rhymes with the empirical formula in Equation (5). Albeit some minor differences in the expression for the $K_{0,NC}$ and the additional attraction term, with the exponent $x = \sin \varphi_c$, we have, rather theoretically, arrived to the same expression as that Mayne and Kulhawy [4] arrived at after investigating the stress state of a significant number of overconsolidated samples. Note however that here the OCR is defined in terms of the apparent isotropic $p_c$ and not of the maximum vertical effective stress experienced by the soil along the $K_0$. It is however possible to write the relationship in terms of the OCR defined by the vertical stress, appendix A.

Now the basic theoretical framework is laid out, some further details can be brought up front.

1. As stated earlier, the condition in Equation (14), that the constant for unloading, $C_N^U$, is the inverse of the constant for loading, $C_N^L$, is not necessary (but it seems natural to assume so and the assumption makes the derivation tidy.) The framework presented above can therefore be considered a special condition. We may generalize it as follows. Suppose $C_l = \frac{1}{C_N^L}$ and $C_u = \frac{1}{C_N^U}$, where $C_N^L$ and $C_N^U$ are as defined in Equation (14), with the condition that $0 < C_l = \frac{1}{N_c} \leq 1$; $C_u \geq 1$ but not necessarily $C_N^U = \frac{1}{C_N^L}$, one obtains:

$$\sigma_h + a = \text{OCR}^{-\frac{(C_u-1)(C_l-1)}{C_u-C_l}} C_l^{-\frac{C_l(C_u-1)}{2(C_u-C_l)}} C_u^{-\frac{C_u(C_l-1)}{2(C_u-C_l)}} (\sigma_v + a). \tag{43}$$

The $K_{0,NC}$ will then be given by

$$K_{0,NC} = C_l^{-\frac{C_l(C_u-1)}{2(C_u-C_l)}} C_u^{-\frac{C_u(C_l-1)}{2(C_u-C_l)}}. \tag{44}$$

Suppose $C_u = fN_c \geq 1$, $C_l = \frac{1}{N_c}$, and the exponent over the OCR is denoted by $x$, one can write

$$K_{0,NC} = N_c^{-\frac{1}{2}} (f)^{-\frac{fN_c(1-N_c)}{2(fN_c^2-1)}}, \tag{45}$$

$$x = \frac{(fN_c - 1)(N_c - 1)}{fN_c^2 - 1}. \tag{46}$$

A special value of $f$ we may be interested to establish is for the condition that the $K_{0,NC}$ is described by Jaky's equation, i.e., $K_{0,NC} = 1 - \sin \varphi_c$. This requires finding the solution of $f$ for the condition

$$f^{\frac{fN_c(N_c-1)}{2(fN_c^2-1)}} = \cos \varphi_c. \tag{47}$$

The closed form solution for $f$ is

$$f = \exp\left\{\frac{-2N_c \ln(\cos \varphi_c)}{N_c - 1} + W\left[\frac{2\ln(\cos \varphi_c)}{N_c(N_c - 1)} \exp\left\{\frac{2N_c \ln(\cos \varphi_c)}{N_c - 1}\right\}\right]\right\}, \tag{48}$$

where W is Lamberts W-function. Equation (48) may be approximated by $f \approx$ 16.586exp{−2.792cos$\varphi_c$}. When $f$ is specified by Equation (48), the exponent, $x$, shows a slight increase over $\sin \varphi_c$. The plots in Figure 6 illustrate how the value of $f$ changes with the *cosine* of the critical state friction angle and, accordingly, how the exponent, $x$, changes with the critical state friction angle. Note that the exponent, $x$, cannot be larger than unity for any value of $f$ in Equation (48). For realistic values of the critical state friction angle, the value of $x$ will be 0.4 - 0.7. For this range, the value of $x$ described by Equations (46) and (48**)** is only about 11% higher than $\sin \varphi_c$.

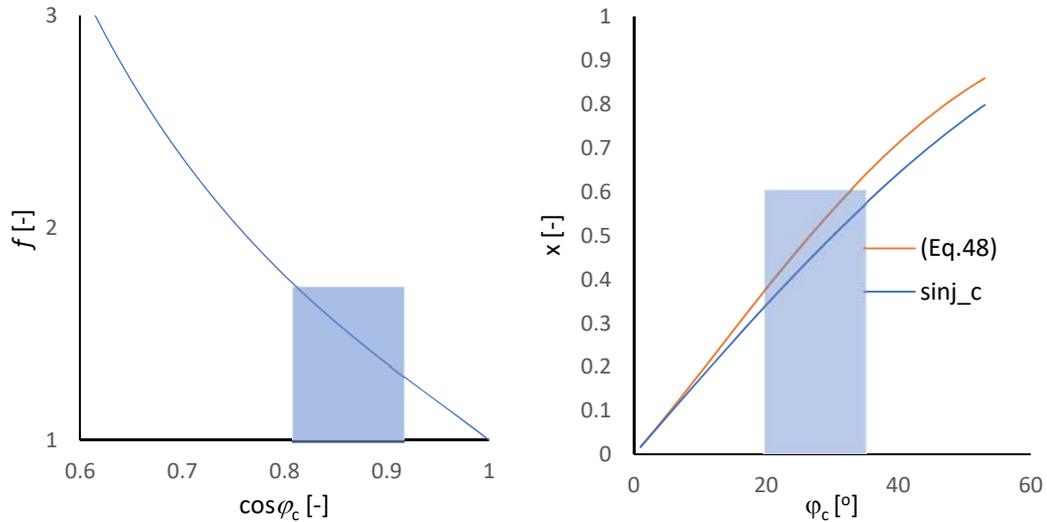

**Figure 6:** Plots of the value of $f$ with the cosine of the critical state friction angle (left) and the exponent $x$ with critical state friction angle (right); shaded regions are for realistic ranges of the critical state friction angle.

2. Note also that the $K_0$ - OCR relationship we so far derived hinges on the assumption that the virgin stress state the soil has ever experienced is created along the $K_{0,NC} = N_c^{-\frac{1}{2}}$ stress path (which may be a naturally preferred condition for maximizing entropy for a given mean effective stress.) Suppose it is created by any other random stress path, say $\sigma_h + a = K_{NC}(\sigma_v + a)$, the corresponding stress proportion that is created by unloading is still exponentially related to the corresponding OCR and is given by

$$\sigma_h + a = K_{NC} \text{OCR}^x (\sigma_v + a). \tag{49}$$

If one wishes to write the stress ratio using an OCR created along any other stress path of proportion $K_{NC} \neq K_{0,NC} = N_c^{-\frac{1}{2}}$ using the $K_{0,NC} = N_c^{-\frac{1}{2}}$, one can still do that but must use some other virtual OCR, say $\text{OCR}_v$ such that

$$\sigma_h + a = K_{0,NC} \text{OCR}_v^x (\sigma_v + a). \tag{50}$$

The $\text{OCR}_v$ may also be multiplicatively split as $\text{OCR}_v = \text{OCR}_r \text{OCR}_i$, where $\text{OCR}_r$ is the real OCR created by any random $K_{NC}$ consolidation path and $\text{OCR}_i$ is the additional OCR that would have been obtained if the mass were normally consolidated along the $K_{0NC} = N_c^{-\frac{1}{2}}$. Suppose $K_{NC} = \beta K_{0,NC}$, then

$$\sigma_h + a = K_{0,NC} \beta \text{OCR}_r^x (\sigma_v + a); \beta = \frac{K_{NC}}{K_{0NC}}. \tag{51}$$

Stress states created by unloading from other $K_{NC}$ stress paths than the $K_{0NC}$ may be the reason that some researchers found higher values for the exponent over the OCR. The comment by Jefferies et al. [16], that there is no unique relation between OCR and $K_0$, may be true for stress states created by unloading along a constrained path along which the soil cannot react in its innate tendencies in one or more of its stress components while some changes occur on one or the other. The writer believes that such stress states are less probable for a natural soil deposit; and therefore, the exponential relationship is applicable for practical purposes.

3. There is some mobilization of the yield function during unloading and, therefore, the assumption that the stress path follows a single yield curve during unloading is not entirely true. However, for shear mobilization towards isotropic stress state, the plastic mobilization of the yield function is assumed to be negligible. This may be studied more generally by employing the theory in an elastoplastic constitutive framework.

4. Whether the constant volume friction angle at the phase transformation fits better instead of the critical state friction angle remains to be investigated. If it happens that the constant volume friction angle at the phase transformation is better fitting, then the $K_0$ will depend on the density of the soil sample.

5. In the theory presented in this paper, the OCR is defined based on unloading. Creep theories suggest that the OCR for clay soils increases with time. It may follow from this that time has some effect on $K_0$. But, recent advancements in this area seem to suggest such may not be true, at least not in the same sense as that of the OCR created by unloading [26]. In the outset, unloading involves a direct change on the stress components while volumetric creep seems to remove the apparent overconsolidation ratio further by creep compaction; and has therefore little to do with the actual change of the stress components except during instances that creep strains are constrained and therefore necessitate stress relaxation. The question would then be, how can, during observation, the OCR that is induced by unloading will be differentiated from that which is induced by time. This aspect needs further investigation. We may, for the sake of brevity and rather less rigorously, suggest a simple approach. The total OCR may be multiplicatively split into that which is caused by unloading, say $OCR_u$, and that which is caused by time, say $OCR_t$, i.e., $OCR = OCR_u \times OCR_t$. Such a split relies on the assumption that one of the components may be determined independently of the other. For instance, the $OCR_t$ may be determined from other assumptions such as the age of the soil deposit. The effect of each on the $K_0$ may then be determined considering the corresponding OCR values. However, there is some fundamental issue that needs to be settled regarding the OCR-age relationship suggested in creep theories. Most creep theories seem to suggest that older deposits would tend to have higher OCR. However, measurements and interpretations of the OCR indicate that the general trend is such that overconsolidation ratio tends towards unity (*i.e.*, towards normally consolidated) with increasing depth. That means, if we follow those theoretical frameworks for creep, then we will be led to the absurd conclusion that deeper soil layers are at a younger age than overlying soil layers unless we admit that the time resistance of soils depends not only on time, as it is usually portrayed, but also on the magnitude of stress. That, time resistance depends on the magnitude of the effective stress state the soil mass is carrying, is something that can be imagined intuitively. But a thorough experimental investigation is yet lacking.

6. Assuming the constant for unloading is the inverse of the constant for loading, the author arrived at an expression for the shear stress at the critical state condition on the maximum mobilized yield curve with an apparent preconsolidation stress, $p_{cl0}$,

$$\tau_{c0} = \frac{1}{2}(N_c-1) N_c^{\frac{1-2N_c}{2(N_c-1)}} OCR^{\frac{N_c \sin\varphi_c}{N_c-1}} (\sigma_v + a). \tag{52}$$

Coincidentally, this is in the same format as that of the SHANSEP [27, 28]; and the undrained shear strength, $c_u$, of clay soils may be estimated using Equation (52). The author intends to treat the full exposition of this and related aspects of the theory in a separate paper.

7. The theoretical framework may be extended to the general stress state considering the generalized CStaD plastic potential function the writer derived [22] based on hypothesis of complementarity of stress-dilatancy conjugates. A preliminary exposition is presented in Appendix B.

# 5 Conclusion

A novel theoretical framework is established for describing the relationships between $K_0$, critical state friction angle and overconsolidation ratio based on the cyclic stress-dilatancy relationship developed by the author. The coefficient of earth pressure at rest for normally consolidated soils is derived based on the assumption that the vertical and the horizontal stress components in a soil deposit tend to be in a proportion that maximizes entropy, in line with Gibbs' equilibrium conditions, unless they are subjected to a loading condition that would force the stress components change their proportion into some other, for instance due to unloading or loading in constrained stress paths. This consideration, for a plane strain condition gave a stress ratio close to the one proposed by Jaky. For soils that are overconsolidated by unloading in shear, a Schmidt type $K_0$-OCR relationship is derived. The author wishes to follow this up with generalizations for other modes of deformation and with investigation of the possible influence non-coaxiality and time may have on the $K_0$.

## Appendix A: K0-OCR relationship for an OCR defined by vertical stresses

Consider the maximum realized vertical stress for the normally consolidated condition be $\sigma_{vc}$. Let the stress state be unloaded along the corresponding unloading potential curve to $(\sigma_h, \sigma_v)$. The vertical overconsolidation ratio is then defined as $\text{OCR}_v = \sigma_{vc}/\sigma_v$. It can be easily shown that

$$\text{OCR} = \text{OCR}_v^{1+\frac{1}{N_c}} \qquad \text{A.1}$$

which implies that

$$\sigma_h + a = K_{0,NC} \text{OCR}_v^{1-\frac{1}{N_c}} (\sigma_v + a). \qquad \text{A.2}$$

Suppose the exponent in the $K_0 - \text{OCR}$ relationship is $m = \sin\varphi_c$, the exponent in the $K_0 - \text{OCR}_v$ relationship is $m_v = \frac{2m}{1+m}$.

## Appendix B: Application of the framework in the *p-q* space

The theory may be extended to a general stress state considering the octahedral effective stress and deviatoric effective stress. For the same, let us consider the generalized CStaD plastic potential function the writer derived [22]

$$q + \bar{p} s C_M^\theta \ln \frac{\bar{p}}{\bar{p}_c} = 0, \qquad \text{A.3}$$

where $q$ is the deviatoric stress, $C_M^\theta$ is stress ratio at the critical state (which depends on the Lode angle $\theta$), $s$ is the CStaD state variable indicating loading and unloading in shear (a stress state mobilizing away from isotropic stress condition is considered loading in shear and $s$ assumes a value of 1, while a stress state mobilizing towards isotropy is considered unloading in shear and in that case $s$ assumes a value of -1), $\bar{p} = p + a$ is the shifted effective octahedral stress in which $a$ is attraction [21] and $\bar{p}_c = p_c + a$ is the shifted apparent pre-consolidation stress (where the plastic potential function intersects the isotropic axis.)

Let a point in a soil volume be normally consolidated to a stress state at A which is along the stress path described by the stress ratio, $M_{NC}$, i.e., $q_A = M_{NC} p_A$, Figure A.1. The same stress state satisfies the intersecting loading and unloading potential pairs at A. Considering the loading potential, one can write

$$q_A = M_{NC} p_A = -C_M^\theta p_A \ln\left(\frac{\bar{p}_A}{\bar{p}_{cl,A}}\right), \qquad \text{A.4}$$

which gives

$$\bar{p}_A = \bar{p}_{cl,A} \exp\left(-\frac{M_{NC}}{C_M^\theta}\right). \qquad \text{A.5}$$

Similarly, considering the unloading potential

$$q_A = M_{NC} \bar{p}_A = C_M^\theta \bar{p}_A \ln\left(\frac{\bar{p}_A}{\bar{p}_{cu,A}}\right) \qquad \text{A.6}$$

gives

$$\bar{p}_A = \bar{p}_{cu,A} \exp\left\{\frac{M_{NC}}{C_M^\theta}\right\}. \qquad \text{A.7}$$

The writer is then led to the following identities:

$$\frac{\bar{p}_{cu,A}}{\bar{p}_{cl,A}} = \exp\left(-2\frac{M_{NC}}{C_M^\theta}\right), \qquad \text{A.8}$$

$$\bar{p}_A^2 = \bar{p}_{cu,A} \bar{p}_{cl,A}. \qquad \text{A.9}$$

The identity $\bar{p}^2/\bar{p}_{cu}\bar{p}_{cl} = 1$ is true for any effective octahedral stress, $p$, contained in the loading and unloading potential pairs with preconsolidation stresses of $\bar{p}_{cl}$ and $\bar{p}_{cu}$. It follows from this that

$$\frac{\bar{p}_A}{\bar{p}_B} = \frac{\bar{p}_{cl,A}}{\bar{p}_{cl,B}} = \frac{\bar{p}_{cl,A}\bar{p}_{cu,B}}{\bar{p}_B^2} = \frac{\bar{p}_{cl,A}\bar{p}_{cu,A}}{\bar{p}_B^2} = \text{OCR}_{AB}. \qquad \text{A.10}$$

Considering unloading from the stress state at A to the stress state at B in the field of the plastic potential curves in Figure A.1, and the fact that $\bar{p}_{cl,A} = \text{OCR}\bar{p}_{cl,B}$, the deviatoric stress at B can be related with the effective octahedral stress at B as

$$q_B = \left\{M_{NC} - \frac{1}{2}C_M^\theta \ln(\text{OCR})\right\} \bar{p}_B, \qquad \text{A.11}$$

or

$$q_B = \ln\left\{\text{OCR}^{-\frac{1}{2}C_M^\theta} \exp\left(M_{NC}\right)\right\} \bar{p}_B, \qquad \text{A.12}$$

or, the stress ratio is given as

$$\eta_B = \left.\frac{q}{\bar{p}}\right|_B = \ln\left\{\text{OCR}^{-\frac{1}{2}C_M^\theta} \exp\left(M_{NC}\right)\right\}. \qquad \text{A.13}$$

In words, the stress ratio created by unloading along one of the unloading plastic potentials from a stress state normally consolidated along a stress path $q = M_{NC}\bar{p}$ is given by $\ln\left(OCR^{-\frac{1}{2}C_M^\theta} e^{M_{NC}}\right)$.

For a yield function described by the apparent pre-consolidation stress, $\bar{p}_{cl,A}$, that is, say OCR times the current apparent preconsolidation stress, $\bar{p}_{cl,B}$, the mean effective stress at the critical state, $\bar{p}_{cs,A}$, is given by:

$$\bar{p}_{cs,A} = \exp\left\{\ln\left(\text{OCR}\,\bar{p}_{cl,B}\right) - 1\right\}. \qquad \text{A.14}$$

The current apparent $\bar{p}_{cl,B}$ can be related to the current effective octahedral stress, $\bar{p}$, as:

$$\ln\left\{\text{OCR}^{-\frac{1}{2}C_M^\theta} \exp\left(M_{NC}\right)\right\} = -C_M^\theta \ln\left(\frac{\bar{p}_B}{\bar{p}_{cl,B}}\right), \qquad \text{A.15}$$

$$\bar{p}_{cl,B} = \text{OCR}^{-\frac{1}{2}} \exp\left(\frac{M_{NC}}{C_M^\theta}\right) \bar{p}_B. \qquad \text{A.16}$$

Substituting Equation A.16 into Equation A.14, the effective octahedral stress at the critical state of the yield function that contains stress state A is given as

$$\bar{p}_{cs,A} = \sqrt{\text{OCR}} \exp\left(\frac{M_{NC}}{C_M^\theta} - 1\right) \bar{p}_B. \qquad \text{A.17}$$

Since the deviatoric stress at the critical state is given by $q_{cs} = C_M^\theta \bar{p}_{cs}$, we are led to:

$$q_{cs,A} = C_M^\theta \sqrt{\text{OCR}} \exp\left(\frac{M_{NC}}{C_M^\theta} - 1\right) \bar{p}_B. \qquad \text{A.18}$$

In general, suppose the soil sample is normally consolidated along the stress path $q = M_{NC}\bar{p}$ to a preconsolidation stress $p_{cl0}$ with a corresponding effective octahedral stress of $p_{cs0}$ at critical state, then the following relations hold for stress states created by following the natural unloading stress paths (i.e., by unloading along the conjugate potential pair to the yield function that contains $p_{cl0}$)

$$\eta = \frac{q}{\bar{p}} = M_{NC} - \frac{1}{2} C_M^\theta \ln \text{OCR} ,$$

A.19

$$\begin{aligned} q_{cso} &= C_M^\theta \sqrt{\text{OCR}} \exp\left(\frac{M_{NC}}{C_M^\theta} - 1\right) \bar{p} \\ &= C_M^\theta \exp\left(\frac{2M_{NC} - \eta}{C_M^\theta} - 1\right) \bar{p} \end{aligned}.$$

A.20

The undrained shear strength, $c_u$, may also be estimated by considering $c_u = \frac{1}{2} q_{cs0}$.

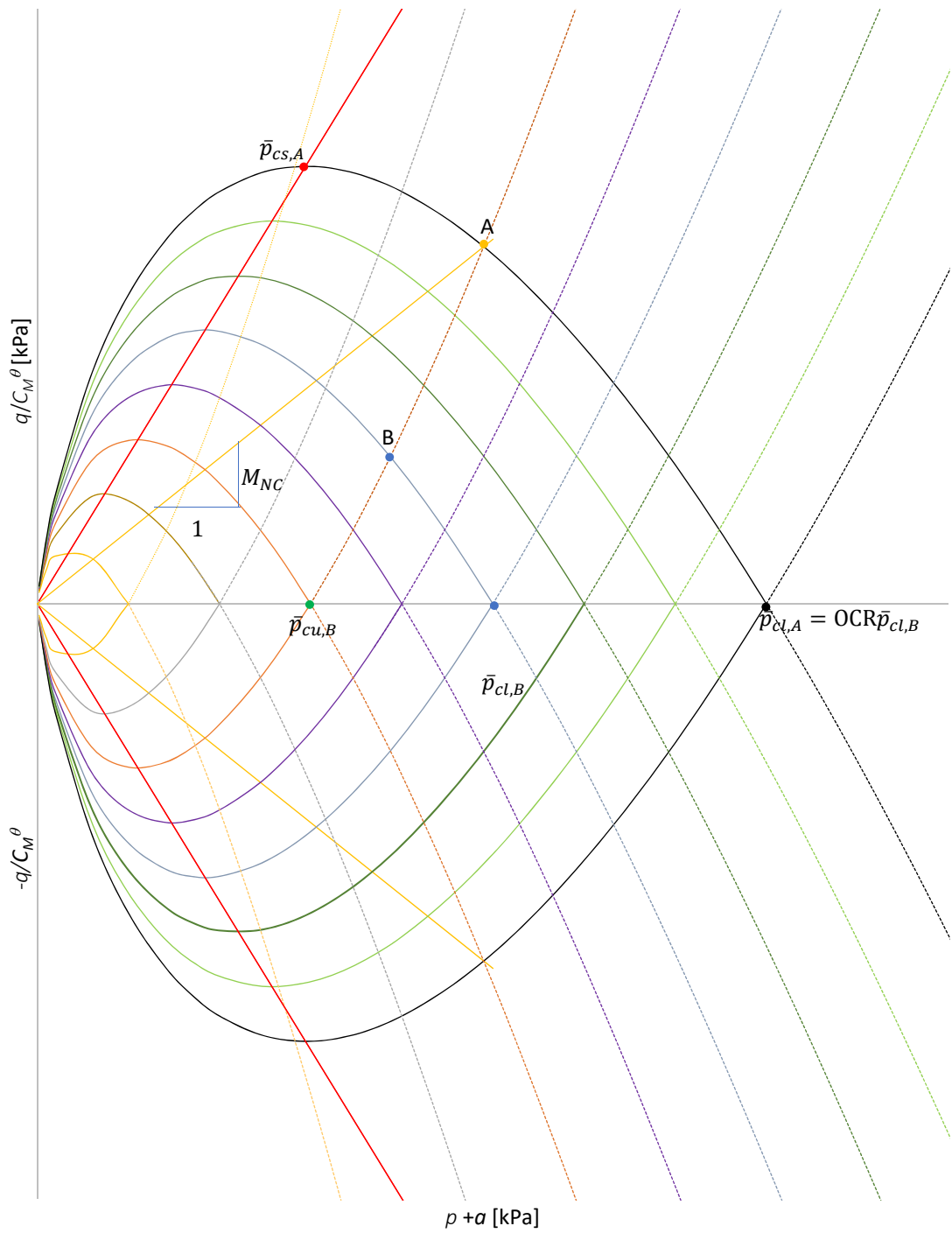

**Figure A.1:** Schematic of the GCStaD loading and unloading potential pairs [22], CS (critical state line)

…on the coefficient of earth pressure at rest